\documentstyle[floats,aps,epsfig,prb]{revtex}

\begin{document}
\draft

\twocolumn[\hsize\textwidth\columnwidth\hsize\csname
@twocolumnfalse\endcsname

\title{\bf Soft tetragonal distortions in ferromagnetic Ni$_2$MnGa and 
related materials from 
first principles}

\author{V.V. Godlevsky and K. M. Rabe}

\address{Department of Physics and Astronomy,\\
Rutgers University,
136 Frelinghuysen Rd, Piscataway, NJ 08854}
\date{\today}
\maketitle

\begin{abstract} 
A detailed examination of the energy landscape, density of states and
magnetic moment of tetragonally distorted ferromagnetic Ni$_2$MnGa was performed using
first-principles local-spin-density (LSD) pseudopotential calculations, varying V as
well as c/a.  The energy of tetragonal Ni$_2$MnGa is found to be {\it nearly constant}
for values of c/a over a wide range, with shallow minima near c/a = 1 and 1.08 in
addition to that near 1.2. This flat energy surface is consistent with the wide range of
observed values of c/a. It also explains the observation of pseudomorphic growth of
Ni$_2$MnGa on GaAs, despite a nominal 3\% lattice mismatch. The related materials
Ni$_2$MnAl, Ni$_2$MnIn and ferromagnetic NiMn were examined for similar behavior, but
all are seen to have a single well-defined minimum at c/a near 1, consistent with
available experimental information.
For NiMn, the ground state antiferromagnetic ordering and structural parameters are correctly predicted within the LSD approach.
\end{abstract}

\pacs{71.15.Nc,81.30.Kf,75.50.-y}

\vskip1pc]

In the design of mechanical actuator devices, both on macroscopic scales and in
thin-film based MEMS, there is continuing interest in identifying and optimizing new
high-performance materials. Materials which exhibit a martensitic transformation and
associated shape memory effect have been shown to be quite useful, though their
high-frequency applications are limited by the slow rate of the martensitic
transformation and poor energy conversion. It has been proposed\cite{james00} that this
deficiency could be addressed by using shape memory materials which are ferromagnetic,
and using applied magnetic fields to control the mechanical response
\cite{james93,james94,ullakko96}.

Therefore, it is of particular interest to identify candidate ferromagnetic martensites
and to understand the transition mechanism, especially the coupling between magnetic and
structural degrees of freedom. The rather small number of known systems includes Fe-Pt,
Co-Ni, Fe-Mn-C, Fe-Ni-C, Fe-Ni-Co-Ti and Fe-Ni, with the Heusler structure compound
Ni$_2$MnGa being the most thoroughly investigated to date 
\cite{ullakko96,ayuela99,webster83,kokorin93,zheludev94,zheludev95,velikohatnyi99}.

In this paper, we investigate the structural energetics of Ni$_2$MnGa and related
materials from first principles, with a particular focus on the tetragonal distortion
associated with the martensitic transition. For comparison, we examine Ni$_2$MnAl and
Ni$_2$MnIn (not observed to be martensitic), and NiMn, related by
replacing Ga with an additional Mn (exhibits a martensitic transition to an
antiferromagnetic tetragonal phase\cite{kasper59}). 
While our calculations are for bulk
crystals, they also allow us to predict the effects of epitaxial stress, which often is
the dominant factor in determining the structure and properties of thin films.

We use the self-consistent pseudopotential plane wave approach\cite{chelikowsky92}
within the local spin density approximation (LSDA). We use
Troullier-Martins\cite{troullier91} pseudopotentials and exchange and correlation
potential in the Perdew-Wang\cite{perdew92} form. The nonlinear core correction
scheme\cite{louie82} was used in the construction of the pseudopotentials. The
pseudopotential cut-off radii are summarized in Table I. The local components of the 
pseudopotentials\cite{kleinman82} are chosen to be l=0 for Ni, Mn, Al and In and l=1 for
Ga. The energy plane-wave cut-off is 70 Ry. The unit cells consist of one formula
unit for Ni$_2$MnX (X = Al, Ga, and In) and two formula units for NiMn. 
The Brillouin zone is sampled by
a 10x10x10 k-point Monkhorst-Pack (MP)\cite{monkhorst76} mesh with zero shift. To be able 
to
calculate the magnetic moment accurately, we do not apply temperature smearing for
the k-point integration. 
In NiMn, we studied two antiferromagnetic phases: 
AF-I, with the Mn atoms with parallel moments located on alternating (1 1 0)
planes; and AF-II, with parallel moments on alternating (1 1 $\bar{1}$)
planes (in the cubic reference system) \cite{sakuma98}.
The 
self-consistent calculations tend to converge to a local ferromagnetic minimum.
To make the system converge to the antiferromagnetic state, approximately 0.2 eV/atom
lower in energy, we constrain the total
magnetic moment to zero.

\begin{table}
\caption{Reference configurations and cut-off radii (a.u.)
used to construct the pseudopotentials.}
\vspace{0.2in}
\begin{tabular}{lddddddd}
\ & $r_s$ & $r_p$ & $r_d$ \\
\hline
\hline
\ Ni $3d^84s^24p^0$ & 2.2 & 2.2 & 2.2  \\
\hline
\ Mn $3d^54s^24p^0$ & 1.9 & 2.6 & 2.0  \\
\hline
\ Al $4s^24p^14d^0$ & 2.3 & 2.3 & 2.3 \\
\hline
\ Ga $4s^24p^14d^0$ & 3.0 & 2.6 & 3.0 \\
\hline
\ In $4s^24p^14d^0$ & 3.0 & 3.0 & 3.0 \\
\end{tabular}
\vspace{0.15in}
\end{table}

For the high temperature cubic structures, our
calculations yield the lattice constants, bulk moduli and magnetic moments given in
Table II. For NiMn, we compare calculated results for a hypothetical ferromagnetic cubic 
phase with 
measurements in the paramagnetic cubic phase.
 The results compare well with experiment and with previous first-principles studies.

\begin{table}
\caption{Lattice constants, bulk moduli and magnetic moments of
Ni$_2$MnX compounds (L2$_1$ structure) and FM-NiMn (B2 structure)
compared to experiment (in parentheses) and previous theoretical
calculations (square brackets), where available.}
\vspace{0.2in}
\begin{tabular}{lddddddd}
\ & a (a.u.)  \\
\hline
\hline
\ FM-NiMn & 5.56 (5.63$^a$) \\
\hline
\ ${\rm Ni_2MnAl}$ & 10.93 (11.01$^b$) [10.98$^d$]\\
\hline
\ ${\rm Ni_2MnGa}$ & 10.91 (11.01$^b$) [10.95$^d$]& \\
\hline
\ ${\rm Ni_2MnIn}$ & 11.32 (11.47$^b$) [11.43$^e$]& \\
\hline
\hline

\ &  B (GPa)  \\
\hline
\hline
\ FM-NiMn & 155  \\
\hline
\ ${\rm Ni_2MnAl}$ & 167 [163$^d$] \\
\hline
\ ${\rm Ni_2MnGa}$ & 170 (146$^c$) [156$^d$] \\
\hline
\ ${\rm Ni_2MnIn}$ & 138 \\
\hline
\hline

\ & $\mu (\mu_B/Mn)$  \\
\hline
\hline
\ FM-NiMn & 4.42 [3.8$^g$] \\
\hline
\ ${\rm Ni_2MnAl}$ & 4.22 (4.19$^b$) [4.03$^d$]\\
\hline
\ ${\rm Ni_2MnGa}$ & 4.22 (4.17$^b$) [4.09$^d$]\\
\hline
\ ${\rm Ni_2MnIn}$ & 4.31 (4.40$^b$) [3.91$^e$] [4.16$^f$]\\
\end{tabular}

$^a$Reference \onlinecite{egorushkin83} \newline
$^b$Reference \onlinecite{webster83} \newline
$^c$Reference \onlinecite{worgull96} \newline
$^d$Reference \onlinecite{ayuela99} \newline
$^e$Reference \onlinecite{daSilva88} \newline
$^f$Reference \onlinecite{kilian00} \newline
$^g$Reference \onlinecite{sakuma98}
\end{table}

Next, we present the results of total energy calculations for uniform tetragonal
distortions of these cubic structures, and discuss them in the context of observed
low-symmetry phases. At least three distinct ferromagnetic low-symmetry phases have been
experimentally observed in bulk Ni$_2$MnGa \cite{kokorin93}, 
two tetragonal ($\beta^\prime$ and $\beta^{\prime\prime\prime}$) and one
orthorhombic ($\beta^{\prime\prime}$).  
The $\beta^\prime$ phase, obtained by cooling of the high-temperature
Heusler (L2$_1$) phase below T$_m \approx$ 200 K, has a tetragonal structure with c/a = 
0.94
(a=5.920 \AA~ and c=5.566 \AA).\cite{webster83} The relative volume change across the
martensitic transition is only 1\%. In addition, there is a shuffle of the (1 1 0)
planes along $[1 \bar{1} 0]$. This has incommensurate periodicity represented
by a wavevector of $\pi /a$[0.43,0.43,0], corresponding to approximately five
interplanar distances \cite{zheludev94,zheludev95}. The transition is preceded by a
cubic premartensitic ($L2_{PM}$) phase below T$_{PM} \approx$ 260 K, characterized by 
softening and
condensation of transverse acoustic phonons $\pi /a$[0.33,0.33,0] with displacements 
similar in
type to the shuffle distortion below T$_m$.  Under uniaxial compression along [1 1 0]
the $\beta^\prime$ phase transforms into the orthorhombic
$\beta^{\prime\prime}$ phase (a = 5.54 \AA, b = 5.78 \AA~ and c = 6.12
\AA)\cite{kokorin93}. X-ray analysis shows that $\beta^{\prime\prime}$
martensite is also modulated with $[1 \bar{1} 0]$ displacements of the (1 1 0) planes
with a periodicity of seven (1 1 0) interplanar distances. The third martensite phase
$\beta^{\prime\prime\prime}$, obtained by further compression along [1 1 0], 
is metastable with respect to the removal of the applied stress \cite{kokorin93}. 
The
$\beta^{\prime\prime\prime}$ structure is a uniform tetragonal distortion of the cubic
Heusler structure (a = 6.44 \AA~and c = 5.52 \AA) with c/a = 1.18. The
relationships between these phases can be summarized as follows:

\begin{equation}
L2_1 \;\; \stackrel{T=260 K}{\longrightarrow} \;\; L2_{PM}
\;\; \stackrel{T=200 K}{\longrightarrow} \;\; \beta^\prime
\;\; \stackrel{[1 1 0]}{\stackrel{stress}{\longrightarrow}}
\;\; \beta^{\prime\prime}
\;\; \stackrel{[1 1 0]}{\stackrel{stress}{\longrightarrow}}
\;\; \beta^{\prime\prime\prime}.
\end{equation}

\begin{figure}
\centerline{\epsfig{file=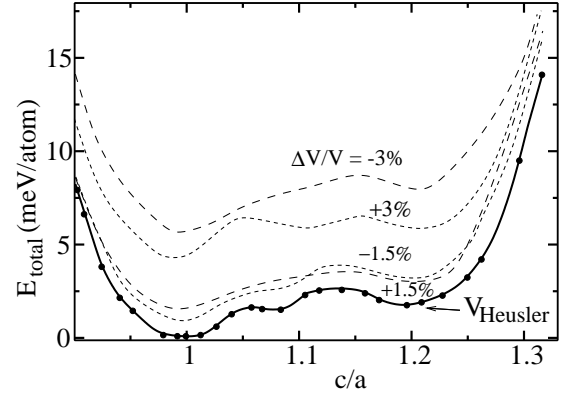,width=3.3in}}
\vspace{0.4cm}
\caption{Total energy as a function of c/a in Ni$_2$MnGa. Each curve
corresponds to a constant volume V per unit cell,
measured relative to V$_{Heusler}$ = 1300 a.u.$^3$
The curves are separated by intervals of $\Delta V/V=1.5$\%.
Each curve represents an interpolation through 26 points, shown explicitly by filled circles for V = V$_{Heusler}$.}
\label{Fig.ni2mnga}
\end{figure}

The calculated energy of Ni$_2$MnGa as a function of c/a for V = V$_{Heusler}$ is shown in
Figure \ref{Fig.ni2mnga}. The region of c/a from 0.9 to 1.3 is examined in considerable
detail, with 26 separate calculations.  Based on the experimental observations of bulk
crystal structures, one would expect to find a local minimum in the energy as a function of
c/a near 1.18 and perhaps another near 0.94. There is a shallow low-energy local minimum
near c/a = 1.19 with total energy 2 meV/atom higher than the L2$_1$ structure.  In the
previous calculations of Ayuela $et.al.$, this minimum is at c/a=1.16 and is slightly lower
in energy than the L2$_1$ structure. The minimum can readily be associated with the pure
tetragonal $\beta^{\prime\prime\prime}$ phase observed by Kokorin $et.al.$ with
c/a=1.18\cite{kokorin93}. In contrast, there is no local minimum or any discernible anomaly
at c/a = 0.94, with a smooth decrease to the minimum at the L2$_1$ structure.  However, our
fine-scale exploration of the energy surface does reveal that the entire energy surface for
0.95 $<$ c/a $<$ 1.25 is remarkably flat, with the total energy varying less than 2.5
meV/atom.  This has significant physical consequences for thin films which will be
discussed further below.

First, we consider the reasons for the absence of a local minimum corresponding to the
ground state $\beta^{\prime}$ phase. It has been suggested that the tetragonal distortion
could be stabilized by a change in volume\cite{ayuela99}. To investigate this hypothesis,
we computed the energy as a function of c/a for varying volumes, with results shown in
Figure \ref{Fig.ni2mnga}. We sampled the volumes in 1.5\% increments of $\Delta V/V$. Even
with the possibility of volume change, the L2$_1$ structure remains the total energy
minimum, with no local minimum corresponding to the $\beta^\prime$ structure.  We remark
that the minimum near c/a = 1.19 is lowered relative to the cubic structure as V increases,
although it never actually becomes more favorable. We have also calculated the total energy
of a pure orthorhombic structure at the experimental lattice constants, which give c/a=1.11
and b/a=1.04.\cite{kokorin93} The calculated energy of this orthorhombic phase is 4 meV per
atom higher than the L2$_1$ phase, consistent with previous calculations.\cite{ayuela99}  
We conclude that
the observed shuffle distortions are crucial to stabilizing the observed $\beta^{\prime}$
and $\beta^{\prime\prime}$ phases.

A number of related materials were examined for comparison. Martensitic transitions have
been observed in the Ni-Mn-Al system, but not in stoichiometric Ni$_2$MnAl
\cite{gejima99,sutou98}. Similarly, there are no reports of low-symmetry phases of
Ni$_2$MnIn. NiMn exhibits a martensitic transition from a cubic paramagnetic phase at T$_m$
= 943 K to a tetragonal phase with c/a = 1.33 which, however, is {\it anti}ferromagnetic
\cite{kasper59,egorushkin83}.

\begin{figure}
\centerline{\epsfig{file=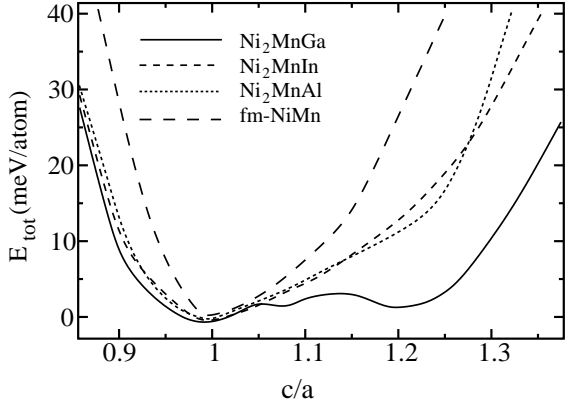,width=3.3in}}
\vspace{0.4cm}
\caption{Total energy as a function of c/a ratio of several related compounds, compared
with Ni$_2$MnGa. The curve for each compound is computed at a constant volume equal to the theoretical V$_{Heusler}$ given in Table II. Each curve represents an interpolation through at least 20 points in the range shown. The energies are given relative to the energy at c/a = 1, with the curves slightly offset for clarity.
}
\label{Fig.related}
\end{figure}

For Ni$_2$MnAl, Ni$_2$MnIn, and ferromagnetic NiMn, the energy as a function of c/a for V =
V$_{Heusler}$ is shown in Figure \ref{Fig.related}.  All three compounds are seen to have a
single well-defined minimum at c/a near 1, consistent with available experimental
information. In the previous calculations of Ayuela $et.al.$, there was another metastable
minimum for Ni$_2$MnAl at c/a=1.22. We find that for Ni$_2$MnAl the shape of this curve is
very sensitive to the k-point sampling density, with a local minimum near c/a=1.2 evolving
into a slope ``discontinuity" for grids denser than 8x8x8.

\begin{figure}
\centerline{\epsfig{file=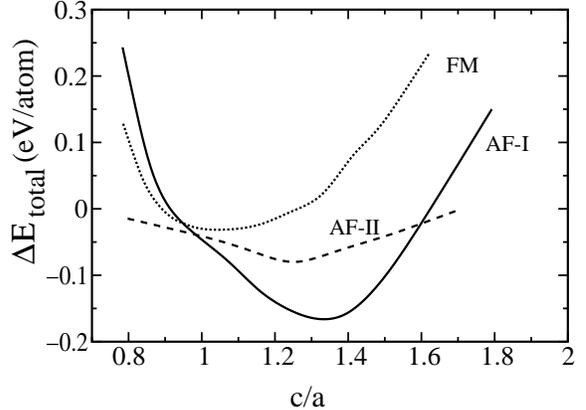,width=3.3in}}
\vspace{0.4cm}
\caption{Total energy of NiMn (FM, AF-I, AF-II phases) as a function of c/a at constant volume.
For FM, AF-I and AF-II phases, we chose the volume of the observed experimental
structure
\cite{kasper59,egorushkin83}, a=3.74 \AA~ and c=3.52 \AA.
}
\label{Fig.nimn}
\end{figure}

\begin{figure}
\centerline{\epsfig{file=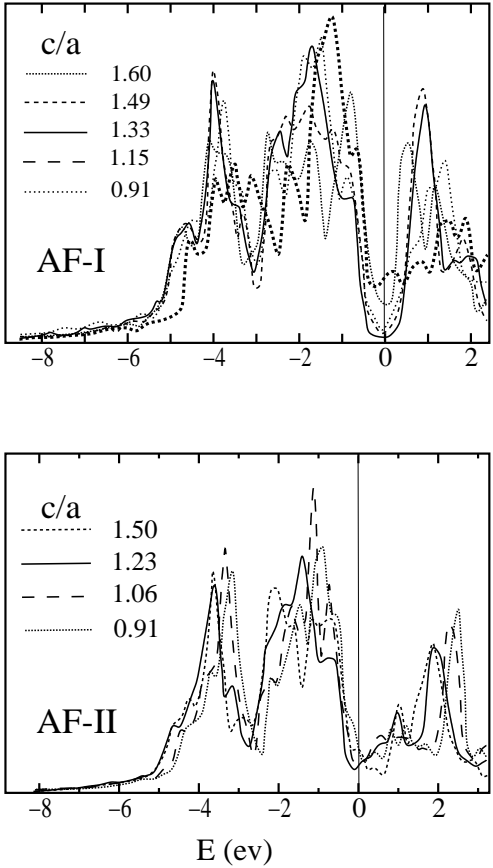,width=3.3in}}
\vspace{0.4cm}
\caption{Density of states for (a) the AF-I and (b) the AF-II phases of NiMn for varying c/a at the same volume as in Figure \ref{Fig.nimn}.
The solid line corresponds
to the minimum energy states for both AF-I and AF-II phases.
}
\label{Fig.NiMnDos}
\end{figure}

When constrained to be ferromagnetic, NiMn has a tetragonal distortion energy curve similar
in shape to that of Ni$_2$MnIn and Ni$_2$MnAl, though with a larger stiffness.  The
substitution of a magnetic atom in the X site has, however, significant consequences for
the magnetic order, stabilizing a type-I antiferromagnetically ordered structure (AF-I) by
190 meV over the lowest-energy ferromagnetic structure (FM). This is in good agreement with
the value of 200 meV obtained in a previous calculation\cite{sakuma98}, performed with
lattice constants fixed at experimental values. As shown in Figure \ref{Fig.nimn}, total
energy minimization for the AF-I NiMn structure leads to a predicted value of c/a = 1.33,
consistent with experiment.\cite{kasper59,egorushkin83} At this value a pseudogap appears
at the Fermi level, as shown in Figure \ref{Fig.NiMnDos}. As the tetragonal distortion
deviates from c/a = 1.33 the pseudogap becomes less pronounced and finally Ni d-states fill
the gap.  The formation of a pseudogap and the associated enhanced stability does not
appear for a distinct antiferromagnetic ordering (AF-II), which is higher in energy by 70
meV and has a much shallower minimum at c/a = 1.24.

\begin{figure}
\centerline{\epsfig{file=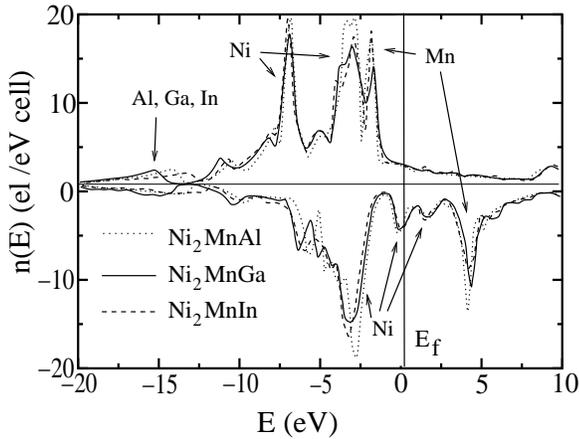,width=3.3in}}
\vspace{0.4cm}
\caption{Density of states in Ni$_2$MnX compounds (X = Al, Ga, In).
The atomic character (obtained by projecting the wave functions onto
individual atoms)
is indicated by appropriate labels.}
\label{Fig.dos}
\end{figure}

The densities of states for cubic Ni$_2$MnX (X = Ga, Al and In) are shown in Figure
\ref{Fig.dos}. For all three compounds, these are quite similar, especially near the Fermi
level.  The total magnetic moment of Ni$_2$MnGa L2$_1$ structure is 4.228 $\mu_B$, which
decomposes into 0.301/3.700/-0.075 $\mu_B$ on Ni/Mn/Ga atoms.  This agrees well with the
previous calculations of References \onlinecite{ayuela99} and \onlinecite{velikohatnyi99}.

\begin{figure}
\centerline{\epsfig{file=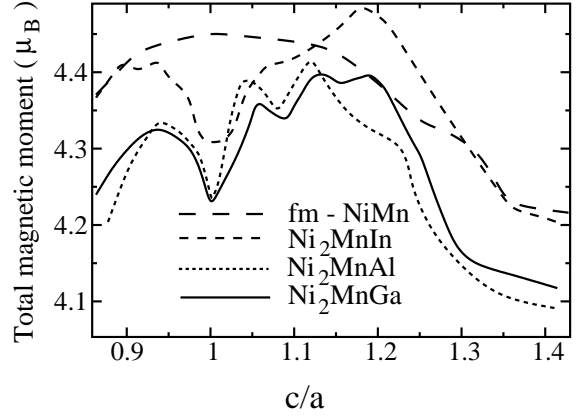,width=3.3in}}
\vspace{0.4cm}
\caption{Total magnetic moments of Ni$_2$MnX (X = Al, Ga, In) and FM-NiMn as a function of c/a, corresponding to the total energy calculations in Figure \ref{Fig.related}.
}
\label{Fig.magmom}
\end{figure}
\begin{figure}
\centerline{\epsfig{file=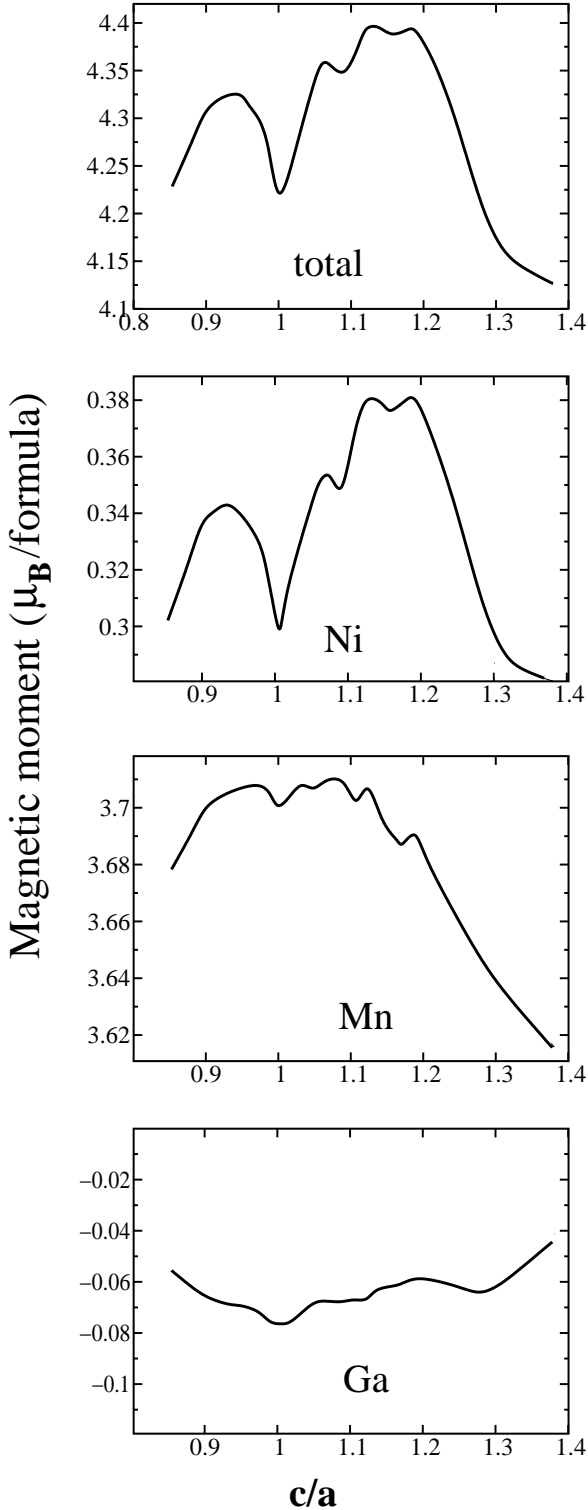,width=3.3in}}
\vspace{0.4cm}
\caption{Total magnetic moment and contributions associated with individual elements
atoms in Ni$_2$MnGa as a function of c/a for V = V$_{Heusler}$. Note the
difference in the vertical scale between the first plot and subsequent plots.
}
\label{Fig.decomp}
\end{figure}

The magnetic moment as a function of c/a, shown in Figure \ref{Fig.magmom}, is again
similar for all three compounds. It shows a sharp minimum at c/a near 1 and and a local
maximum at c/a $\approx$ 0.94. From the atomic decomposition of this curve for Ni$_2$MnGa
in Figure \ref{Fig.decomp}, we see that the features can be directly attributed to the Ni
contribution, most likely associated with a small Ni-derived peak in the minority density
of states located just below the Fermi level. It is interesting to note that the position
of this maximum coincides with the value of c/a for the $\beta^\prime$ martensitic
structure.

The softness of the tetragonal distortion in Ni$_2$MnGa has significant implications for
the growth and properties of thin films.  Efforts to grow single-crystal thin films on a
GaAs substrate (with 3.1\% lattice mismatch) have been carried out by MBE
\cite{dong99,dong00}. Using a Sc$_{0.3}$Er$_{0.7}$As interlayer, a tetragonal-structure
film with lattice parameters a = 0.565 nm and c = 0.612 nm was obtained.  This is exactly
the structure we expect based on the total energy curves in Fig. \ref{Fig.ni2mnga}.  As the
in-plane a is reduced to 0.565 nm to match the substrate, the c axis increases to maintain
the volume of the cubic Heusler phase, which minimizes the total energy over the whole
range of c/a considered. The energy of the resulting structure, at c/a = 1.08, is less than
2 meV/atom higher than the cubic structure, explaining the observation that the thickness
of this pseudomorphic film, 300 \AA, far exceeds the critical thickness expected for the
given lattice mismatch.  While the large epitaxial stress should affect the structure and
properties significantly, if the film can be released it should exhibit a martensitic
transition and shape-memory properties.

In another recent experiment\cite{dong00} a 450 \AA~thick film of Ni$_2$MnGa was grown by
MBE on a GaAs substrate using a NiGa interlayer. The structure of this film was reported to
be tetragonal, with c/a = 1.05 (a=5.79 \AA~and c=6.07 \AA). In this case, the lattice
constant of Ni$_2$MnGa is a close (0.6\%) match to the interlayer, leading to the
expectation that the structure of the film should be cubic. Instead, the c parameter is
expanded, resulting in a volume 3\% larger than the volume of the L2$_1$ structure. One
explanation might be a slight change in the stoichiometry of the film.
Modelling the effect of this shift as an negative applied hydrostatic pressure, we now
focus on the $\Delta V/V$ = +3\% curve in \ref{Fig.ni2mnga}. The flatness of this curve is
similar to that for V = V$_{Heusler}$. Thus, although the expanded system is formally
mismatched to the NiGa interlayer, the constraint a = 5.79 \AA~is associated with an energy
cost of less than 2 meV/atom, which would allow pseudomorphic growth of this tetragonal
film far beyond the critical thickness.
 
This study highlights the fact that the energy surface of Ni$_2$MnGa is far from simple.  
Studies of additional instabilities of the cubic Heusler structure via density-functional
perturbation theory are in progress. In particular, the energies of shuffle distortions as a
function of wavevector can be obtained with this method for a better understanding of the
observed $\beta^\prime$ ground state structure. This information will also allow us to model the
properties of the high-temperature cubic Heusler phase within an effective Hamiltonian approach.
We expect that these properties reflect the presence of large local distortions around the
average cubic structure and there should be interesting differences from the properties of other
Heusler phases whose ground state structure is cubic.

In conclusion, we performed first principles calculations of the total energy as a function of
tetragonal distortion (c/a) and volume V for Ni$_2$MnGa and related compounds Ni$_2$MnAl,
Ni$_2$MnIn, and NiMn. The total energy of Ni$_2$MnGa at constant volume is remarkably flat in the
range 0.95$<$c/a$<$1.25, varying less than 2.5 meV/atom. This provides an explanation of the
surprisingly high compliance of single-crystal films grown by MBE. In contrast, the related
materials Ni$_2$MnAl, Ni$_2$MnIn, and ferromagnetic NiMn have a single well-defined minimum near
c/a = 1. The densities of states and c/a dependence of the magnetic moments are quite similar in
all compounds studied, leading us to conclude that the unique behavior of Ni$_2$MnGa is the
result of fine tuning. It may be possible to achieve this in other systems through appropriate
alloying.

We thank R. D. James, K. Bhattacharya, C.J. Palmstr$\o$m, J. Dong, C. Bungaro, X. Huang
and A. Ayuela for valuable discussions. 
This work was supported by AFOSR/MURI F49620-98-1-0433.

\end{document}